\documentclass[12pt]{article}
\usepackage{pic04}
\usepackage{hyperref}
\usepackage{url}
\usepackage{graphicx}
\usepackage{array}
\usepackage{amssymb}
\usepackage{amsmath}
\usepackage{hhline}

\begin{document}

\title{\bf CP Violation in ${\bf B}$ Decays and the CKM Matrix} 

\author{
Emmanuel Olaiya        \\
{\em Rutherford Appleton Laboratory} \\
{\em Chilton,Didcot,Oxon,OX11 0QX,UK}}
\maketitle

%
%
%
%
%
%
\vspace{4.5cm}
%

\baselineskip=14.5pt
\begin{abstract}
Results by the BaBar and Belle experiments of the measurements $\sin 2\alpha$, $\sin 2\beta$ and  $\sin (2\beta + \gamma)$ along with the CKM matrix elements $V_{cb}$ and $V_{ub}$ are presented. The BaBar and Belle experiments $b \rightarrow sc\bar{c}$ results for $\sin 2\beta$ are in good agreement with each other, comfortably establishing $CP$ violation within $B$ decays. However, there is a 3.5 standard deviation between this result and $\sin 2\beta$ measured by the Belle experiment using  $B^0 \rightarrow \phi K^0_S$ decays. Belle also find evidence for $CP$ violation through time dependent measurements of the decay  $B^0 \rightarrow \pi^+ \pi^-$, whilst a tighter constraint has been placed on the unitarity angle $\alpha$ by BaBar, using time dependent studies of the decay $B^0 \rightarrow \rho^+\rho^-$ .

\end{abstract}
\newpage

\baselineskip=17pt

\section{Introduction}

Since the discovery of $CP$ violation in 1964~\cite{fitch}, great effort has been placed in understanding the origin and mechanism of this asymmetry. In 1973, Kobayashi and Maskawa proposed a model where $CP$ violation is accommodated within the weak interaction as an irreducible complex phase in the quark mixing matrix~\cite{km}. Today we refer to this as the CKM matrix and with three known fermion families it can be written as

\begin{equation}V =
\left(\begin{matrix}
V_{ud} &  V_{us} & V_{ub}\\
V_{cd} &  V_{cs} & V_{cb}\\
V_{td} &  V_{ts} & V_{tb}
\end{matrix}\right)
\end{equation}

\noindent The unitary nature of the matrix implies there are six orthogonality
conditions obtained by row-column multiplication with its inverse. One of these conditions which is relevant to $B$ decays is:

\begin{equation}
V_{ud}V^*_{ub} + V_{cd}V^*_{cb} + V_{td}V^*_{tb} = 0. \hspace*{1.5cm}\label{buni}
\end{equation}

\noindent Each orthogonality condition requires the sum of three complex numbers to be zero,
and therefore can be represented geometrically in the complex plane as a triangle. These are
known as `unitary triangles', and for the condition expressed in Equation~\ref{buni}, we refer to the angles in the triangle as $\alpha$ = $\phi_2$, $\beta$ = $\phi_1$ and $\gamma$ = $\phi_2$.

Two $B$-factories have been built, BaBar at SLAC USA and Belle at KEK Japan, with the aim of measuring $CP$ violation in the $B$ system and to constrain further the CKM matrix using information from $B$ meson decays.

\section{Time evolution of $B$ decays and CP violation}\label{tevo}

We can write the time dependent amplitude of the $B^0$ state at $t$ = 0 decaying into a final state $f$ as 

\begin{equation}
\langle f|H|B^0(t) \rangle =  e^{-imt}e^{\Gamma t/2}\biggl[A_f\cos\frac{1}{2}\Delta mt + i\frac{q}{t}\overline{A}_f\sin \frac{1}{2}\Delta mt\biggr],
\end{equation}

\noindent where $A_f$ = $\langle f|H|B^0(t) \rangle$ is the amplitude of the $B^0$ decay to the final state $f$, $\overline{A}_f$ is the amplitude for $\overline{B}^0$ and $p,q$ give the weak eigenstates $B^0_{L,H}$ in the ($B^0$, $\overline{B}^0$) basis $|B^0_{L,H}\rangle = p|B^0\rangle \pm q|\overline{B}^0\rangle$. The average between the H and L masses is $m$ and the difference is $\Delta m$,  whilst we make the approximation $\Gamma_H = \Gamma_L = \Gamma$. When both  $B^0$ and $\overline{B}^0$ can decay to the same final state $f$, $CP$ violation can occur through the interference between mixing ($q/p$) and the decay ($\overline{A}_f/A_f$), even if $CP$ is conserved in both $|q/p| = |\overline{A}_f/A_f| = 1$. Pairs of $B$ mesons are generated at the $\Upsilon (4S)$ resonance where the two mesons oscillate coherently between the  $B^0$ and $\overline{B}^0$ until one decays. We consider the time interval $\Delta t$ between the decay of one $B$ flavor eigenstate (or ``tag'') and the decay of the other $B$ to a $CP$ eigenstate $f$ where the decay rate in terms of $\Delta t$ can be written as

\begin{equation}
\frac{d \Gamma^f_\pm (\Delta t)}{d \Delta t} \propto e^{-|\Delta t|/\tau}(1 \pm {\cal I}m\lambda_f\sin (\Delta m \Delta t)),
\end{equation}

\noindent where $\lambda_f = q\overline{A}_f/pA_f$ and it is assumed $|\lambda_f| = 1$ and the $+(-)$ sign represents the $B^0$($\overline{B}^0$) tag. In the $\Upsilon (4S)$ center of mass frame the $B$ meson is nearly at rest, leading to a strong correlation between the reconstructed mass and the missing mass of the partner $B$. Therefore the typical choice for an independent pair of kinematic variables is

\begin{equation}
\Delta E = E^*_{beam},  \hspace{1cm}   m_{ES} = \sqrt{E^{*2}_{beam} - |P^*_B|^2},
\end{equation}

\noindent where the asterisk refers to the $\Upsilon (4S)$ frame and the subscript $B$ denotes the reconstructed $B$. True $B$ decays tend to be peak at $\Delta E$ = 0 and $m_{ES}$ = $m_B$. The tagged $B$ does not need to be fully reconstructed as the important information required is simply the decay vertex and whether it's a $B^0$ or $\overline{B}^0$. The flavor of the tagged $B$ is determined by the sign of the charge in the  ($B^0 \rightarrow \ell^+,\overline{B}^0 \rightarrow \ell^-,B^0 \rightarrow K^+, \overline{B}^0 \rightarrow K^-$), leading charged track. The efficiency $\epsilon$ and the mistag fraction $w$ for each tagging algorithm is measured with fully reconstructed eigenstate decays. The effective efficiency is given by $Q = \epsilon(1 - 2w)^2$, where the BaBar and Belle experiments report average measured values of ($28.1 \pm 0.7$)\% and   ($28.8 \pm 0.6$)\% respectively.

The decay rate as a function of $\Delta t$ is modified as a consequence of the tagging efficiency and also the convolution with the resolution function ${\cal R}$. The resulting decay rate becomes

\begin{equation}\label{eq:tevo}
\frac{1}{\Gamma^f}\frac{d \Gamma^f_\pm (\Delta t)}{d \Delta t} = \frac{ e^{-|\Delta t|/\tau}}{4\tau}(1 \pm (1 -2w){\cal I}m\lambda_f\sin(\Delta m \Delta t)) \otimes {\cal R},
\end{equation}

\noindent where $\tau$ is the $B$ lifetime. We look for asymmetries between $B^0$ and $\overline{B}^0$  tagged events having a $\sin(\Delta m\Delta t$) dependence with a known angular frequency $\Delta m$ and an amplitude that is given by ${\cal I}m\lambda_f$ multiplied by the dilution factor $(1 -2w)$. The interesting physics lies within ${\cal I}m\lambda_f$ as it contains the factor $q/p$ which is common to all decay modes. Depending on the decay mode under study, it is ${\cal I}m\lambda_f$ which is measured to help provide information on $\alpha$, $\beta$ or $2\beta + \gamma$.

\section{sin2$\alpha$ from $B \rightarrow \rho^+ \rho^-$ and $B \rightarrow \pi^+ \pi^-$ }
  
Considering decays $B^0 \rightarrow hh$ ($B \rightarrow \rho^+ \rho^-$ or $B \rightarrow \pi^+ \pi^-$), the time dependent decay rate can be further written as

\begin{equation}
\frac{ e^{-|\Delta t|/\tau}}{4\tau}\biggl[1 \pm \{S_{hh}\sin (\Delta m \Delta t) - C_{hh}\cos (\Delta m \Delta t)\} \biggr],
\end{equation}

\noindent where the $CP$ violating asymmetry parameters ${\cal S}_{hh}$ and ${\cal C}_{hh}$\footnote{These parameters $S_{hh}$ and $C_{hh}$ will be used throughout the document with subscripts referring to the decay in question.} (= $-{\cal A}_{hh}$)  are expressed as  ${\cal C}_{hh}$ = (1  $- |\lambda_{hh}|^2$)/(1 + $|\lambda_{hh}|^2$) and ${\cal S}_{hh}$ = 2Im$\lambda_{hh}$/(1 + $|\lambda_{hh}|^2$), where $\lambda_{hh}$ is a parameter that depends on $B^0 - \overline{B^0}$ mixing and the amplitudes for  $B^0$ and $\overline{B^0}$ decaying to $hh$. If these decays proceed only via $b \rightarrow u$ tree amplitudes, ${\cal S}_{hh} = \sin 2\alpha$ and   ${\cal C}_{hh} = 0$. If there are penguin contributions to the amplitude then  ${\cal S}_{hh}$ = $\sqrt{1 -  {\cal C}_{hh}^2}\sin 2\alpha_{eff}$ and ${\cal C}_{hh} \neq 0$ . A limit on the difference between $\alpha$ and $\alpha_{eff}$ can be set using the Grossman-Quinn bound~\cite{grossquinn} which is

\begin{equation}
\cos(2\alpha - 2\alpha_{eff}) \geq  1 - \frac{2B^{00}}{B^{+0}},
\end{equation}

\noindent where $B^{00}$ is the branching fraction of $B^{0} \rightarrow h^0h^0$ and  $B^{+0}$ is the branching fraction of  $B^{+} \rightarrow h^+h^0$.

Based on a dataset of 81~fb$^{-1}$ and 140~fb$^{-1}$ used by the BaBar and Belle experiments respectively, the results in Table~\ref{tab:pipi} where obtained for the decay $B^0 \rightarrow \pi^+ \pi^-$.

\begin{figure}[!htbp]
\begin{center}
\scalebox{0.70}{%
 \includegraphics[width=.59\linewidth]{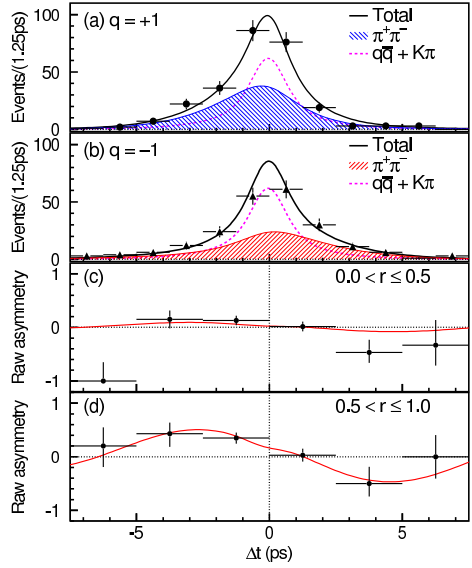}
 }
\caption{{\it The Belle $\Delta t$ distribution for 483 $B^0 \rightarrow \pi^+ \pi^-$ candidates; (a) 264 candidates tagged as $B^0$ decays; (b) 219 candidates tagged as $\overline B^0$; (c) The asymmetry for $0 < r \leq 0.5$ and (d) with $0.5 < r \leq 1.0$.The solid line shows the result of the unbinned maximum likelihood fit to the $\Delta t$ distributions of 1529 $B^0 \rightarrow \pi^+ \pi^-$ candidates. }}
\label{fig:pipidt}
\end{center}
\end{figure}

\begin{table}[!ht]
\begin{center}
\begin{tabular}{|c|c|c|}
\hline
\multicolumn{1}{|c|}{\raisebox{-1.5ex}[0cm][0cm]{Measurements}}&
\multicolumn{2}{c|}{{Experiment}}\\
\hhline{|~--|}
&
BaBar&
Belle\\
\hline
${\cal S}_{\pi^+\pi^-}$&
-0.40  $\pm$ 0.22  $\pm$ 0.03 &
-1.00 $\pm$ 0.21 $\pm$ 0.07\\
${\cal C}_{\pi^+\pi^-}$&
-0.19 $\pm$ 0.19 $\pm$ 0.05&
 -0.58 $\pm$ 0.15 $\pm$ 0.07\\
$|\alpha$ - $\alpha_{eff}|$ &
$< 48^\circ$ @ 90\% CL&
 $-$ \\
\hline
\end{tabular}
\end{center}
\caption{{\it BaBar and Belle measurements of the $CP$ violating parameters ${\cal S}_{\pi^+\pi^-}$ and ${\cal C}_{\pi^+\pi^-}$}.}\label{tab:pipi}
\end{table}

BaBar measure values of  ${\cal S}_{\pi^+\pi^-}$ and ${\cal C}_{\pi^+\pi^-}$ that are consistent with 0 and  no evidence of $CP$ violation. Belle however measure a value of ${\cal S}_{\pi^+\pi^-}$ that is 5.2 standard deviations from 0 with a claim to observation  of $CP$ violation due to interference through the mixing and ${\cal C}_{\pi^+\pi^-}$ that is 3.2 standard deviations from 0 with evidence of direct $CP$ violation. The $\Delta t$ distribution is shown in Figure~\ref{fig:pipidt}.

The BaBar experiment has observed the decay $B^0 \rightarrow \rho^+ \rho^-$ and using branching fraction measurements for $B^0 \rightarrow \rho^0 \rho^0$ and $B^+ \rightarrow \rho^+ \rho^0$ they find that $|\alpha - \alpha_{eff}|$ $\leq$ $12.9^\circ$ at a 68\% confidence level for this mode, providing more sensitivity to the measurement of $\alpha$. To be considered is the fact that  $B^0 \rightarrow \rho^+ \rho^-$ is a vector$-$vector decay which can proceed via 3 helicity amplitudes. The decay can be longitudinally polarized ($\lambda$ = 0) and be a pure $CP$ eigenstate or it can be transversely polarized ($\lambda = \pm 1$) and be a mix of even and odd eigenstates. The polarization measured by BaBar was found to be mainly longitudinal, contributing a fraction $f_{long}$ = 0.99 $\pm$ 0.03 $\pm$ 0.03 of the total decay. Analysing the time dependent amplitude of the decay $B^0 \rightarrow \rho^+\rho^-$ the following results were obtained:

\begin{center}
${\cal S}_{\rho^+\rho^-}$ = $-0.19$ $\pm$ 0.33 $\pm$ 0.11\\
${\cal C}_{\rho^+\rho^-}$ = $-0.23$ $\pm$ 0.24 $\pm$ 0.14\\
\end{center}

\begin{figure}[!htbp]
\begin{center}
\scalebox{0.8}{%
 \includegraphics[width=.59\linewidth]{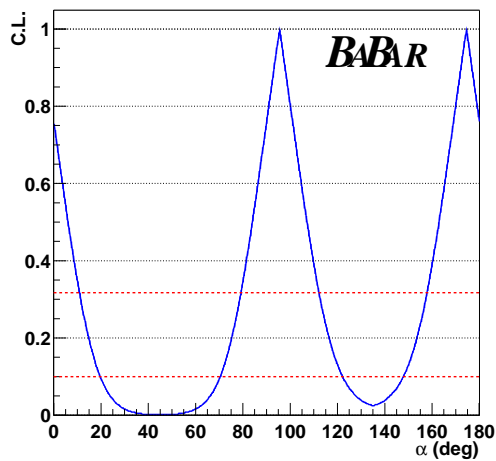}
 }
\caption{\label{fig:alphascan}{\it Isospin analysis confidence level scan or the CKM angle $\alpha$. The input variables are ${\cal B}(B^0 \rightarrow \rho^+ \rho^-$), $f_L$, ${\cal S}_{\rho^+\rho^-}$, ${\cal S}_{\rho^+\rho^-}$  and  ${\cal B}(B^0 \rightarrow \rho^+ \rho^0$)}}
\end{center}
\end{figure}

With the information from  $B^0 \rightarrow \rho^+ \rho^-$ a new value of $\alpha$ is obtained from a CKM fit shown in Figure~\ref{fig:alphascan} corresponding to $\alpha$ = $96^\circ$ $\pm$ $10^\circ_{stat}$ $\pm$ $4^\circ_{syst}$ $\pm$ $13^\circ_{penguin}$, where the first error is statistical, the second is systematic and the third is due to the uncertainty from penguin contributions.

\section{$\sin 2\beta$ from charmonium modes}\label{sec:sin2b}

\begin{figure}[!htbp]
\begin{center}
 \scalebox{.8}{
  (a) \includegraphics[width=.5\linewidth,bb=132 650 335 736]%
   {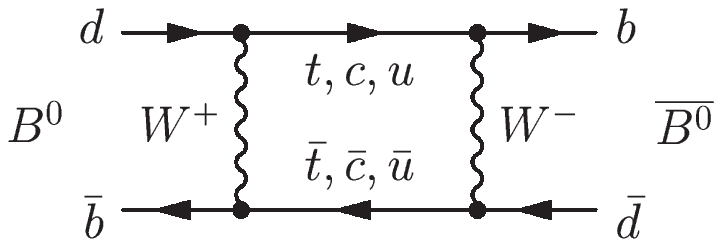}\qquad
  (b) \includegraphics[width=.39\linewidth,bb=132 650 294 732]%
   {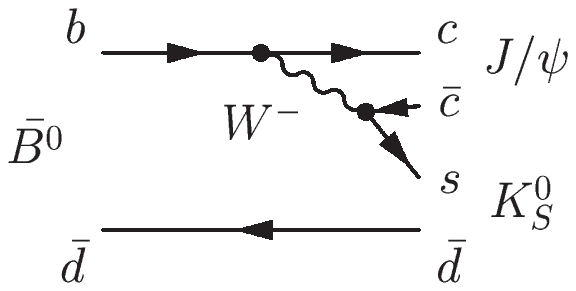}
 }
\end{center}
\vspace{-7mm}
 \caption{\label{fig:feyndiags}{\it (a)  $B$-mixing box-diagram; (b)
dominant diagram for $B \rightarrow c\bar{c}K^0_S$ decays.}}
\end{figure}

For charmonium $K$ decays the relationship between the factor $q/p$ discussed in Section~\ref{tevo} and elements of the CKM matrix can be calculated from Figure~\ref{fig:feyndiags}(a). The loop in the box diagram is dominated by the virtual $t$ quark, since its large mass is responsible for violating the GIM mechanism that would otherwise suppress the mixing. One finds

\begin{equation}
\frac{q}{p} = \frac{V^*_{tb}V_{td}}{V_{tb}V^*_{td}},
\end{equation} 

\noindent which is equal to $e^{-2i\beta}$ in the Wolfenstein phase convention. This information can be used to calculate the amplitude of the sine term in Equation~\ref{tevo}. From Figure~\ref{fig:feyndiags}(b) we see that the ratio of amplitudes for charmonium $K$ decays is

\begin{equation}
\frac{\overline{A}_f}{A_f} = \eta_f\biggl(\frac{V_{cb}V^*_{cs}}{V^*_{cb}V_{cs}}\biggr)\biggl(\frac{p}{q}\biggr) = \eta_f\biggl(\frac{V_{cb}V^*_{cs}}{V^*_{cb}V_{cs}}\biggr)\biggl(\frac{p}{q}\biggr)\biggl(\frac{V^*_{cd}V_{cs}}{V_{cd}V^*_{cs}}\biggr)\biggl(\frac{p}{q}\biggr) = \eta_f,
\end{equation}
 
\noindent therefore $\lambda_f = e^{-2i\beta}$ and ${\cal I}m\lambda_f$ = $-\eta_f\sin 2\beta$.

\begin{figure}[htbp]
\scalebox{1}[.9]{%
 \includegraphics[width=.49\linewidth]{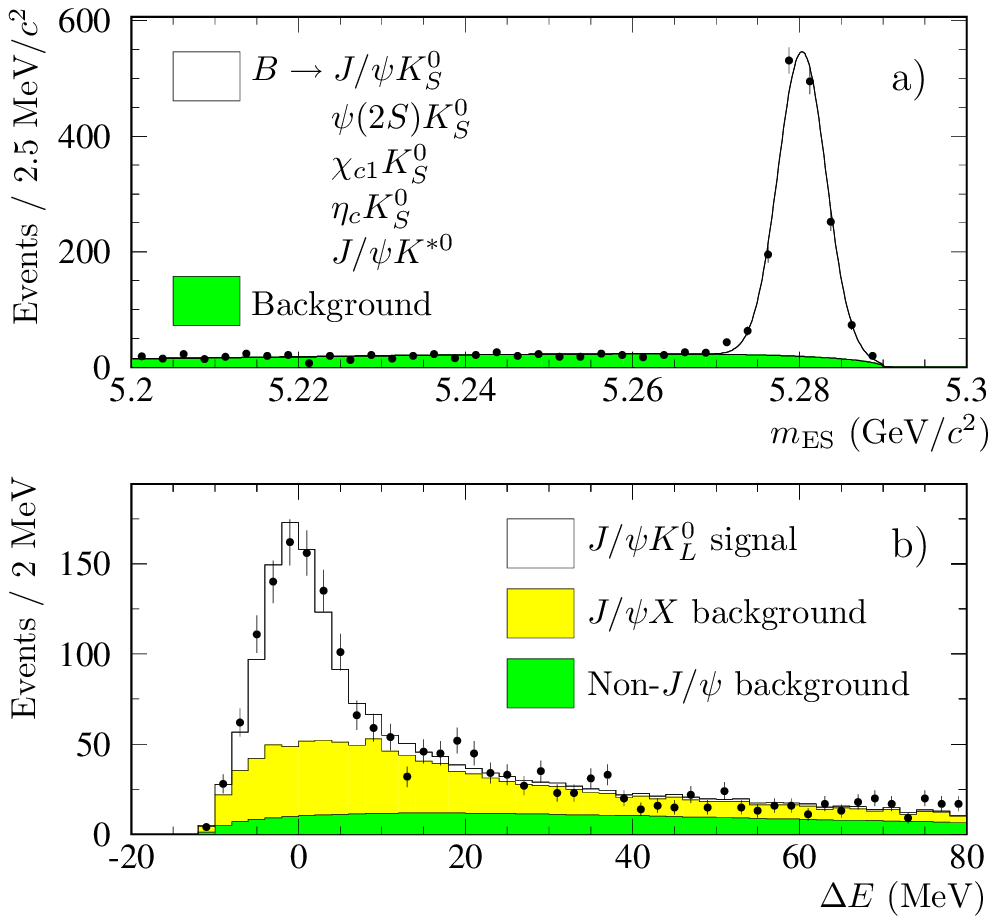}\quad
 \includegraphics[width=.48\linewidth]{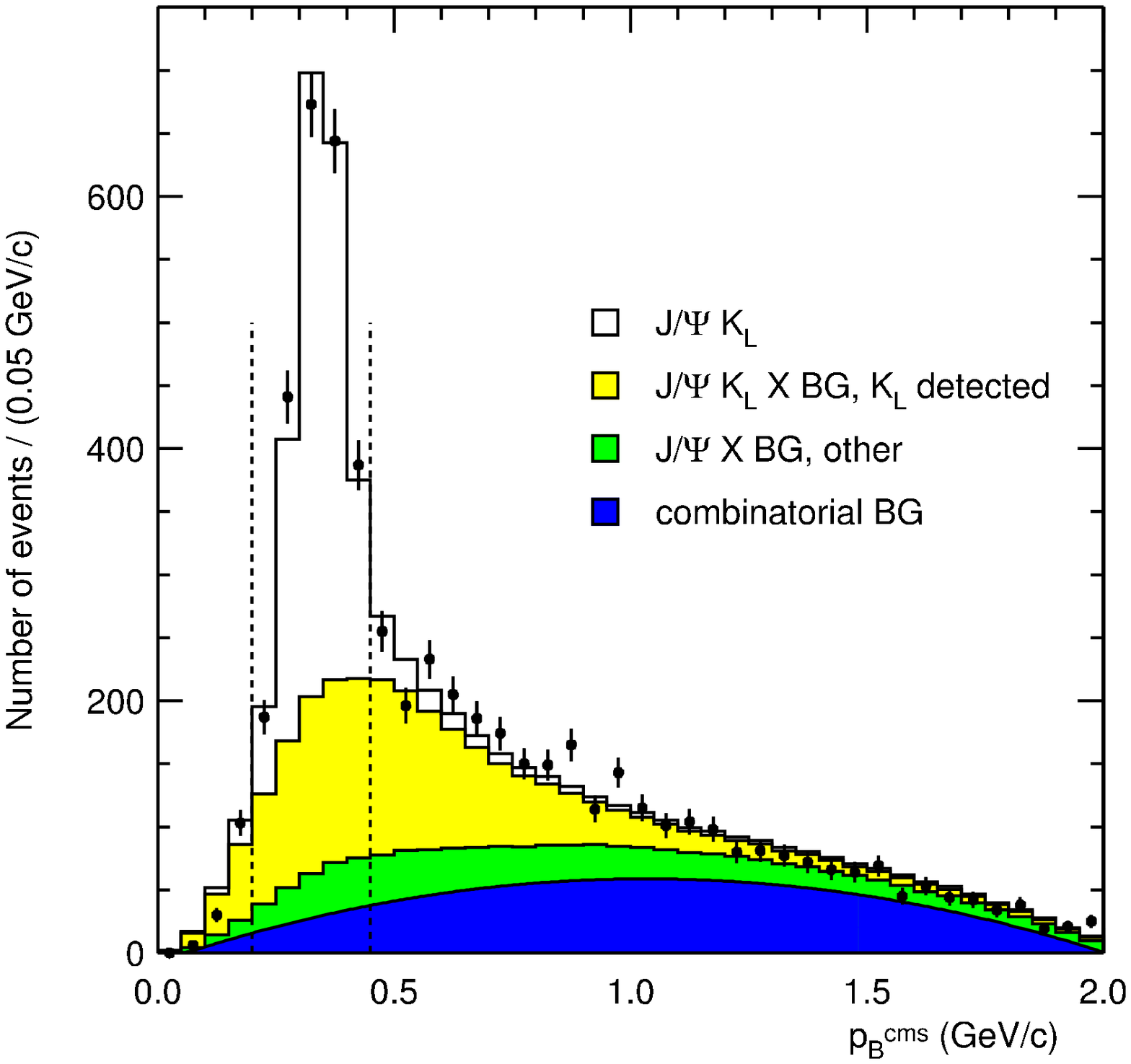}
 }
\caption{\label{fig:ccKdistros}{\it Distributions of $B \rightarrow c\bar{c}K$ events  from BaBar (left) and Belle (right), selected with flavor tags. The top left plot  shows the $m_{ES}$ distributions of the $CP$ odd $c\bar{c}K^0_S$ modes whilst the the bottom left distribution shows $\Delta E$ for the the $CP$ even $c\bar{c}K^0_L$ mode. The right plot shows the momentum ($P_B^{CMS}$) of the reconstructed $CP$ even $B$ candidates in the $\Upsilon(4S)$ center of mass frame}}
\end{figure}

\begin{figure}
 \scalebox{1}[0.8]{\includegraphics[width=.505\linewidth]{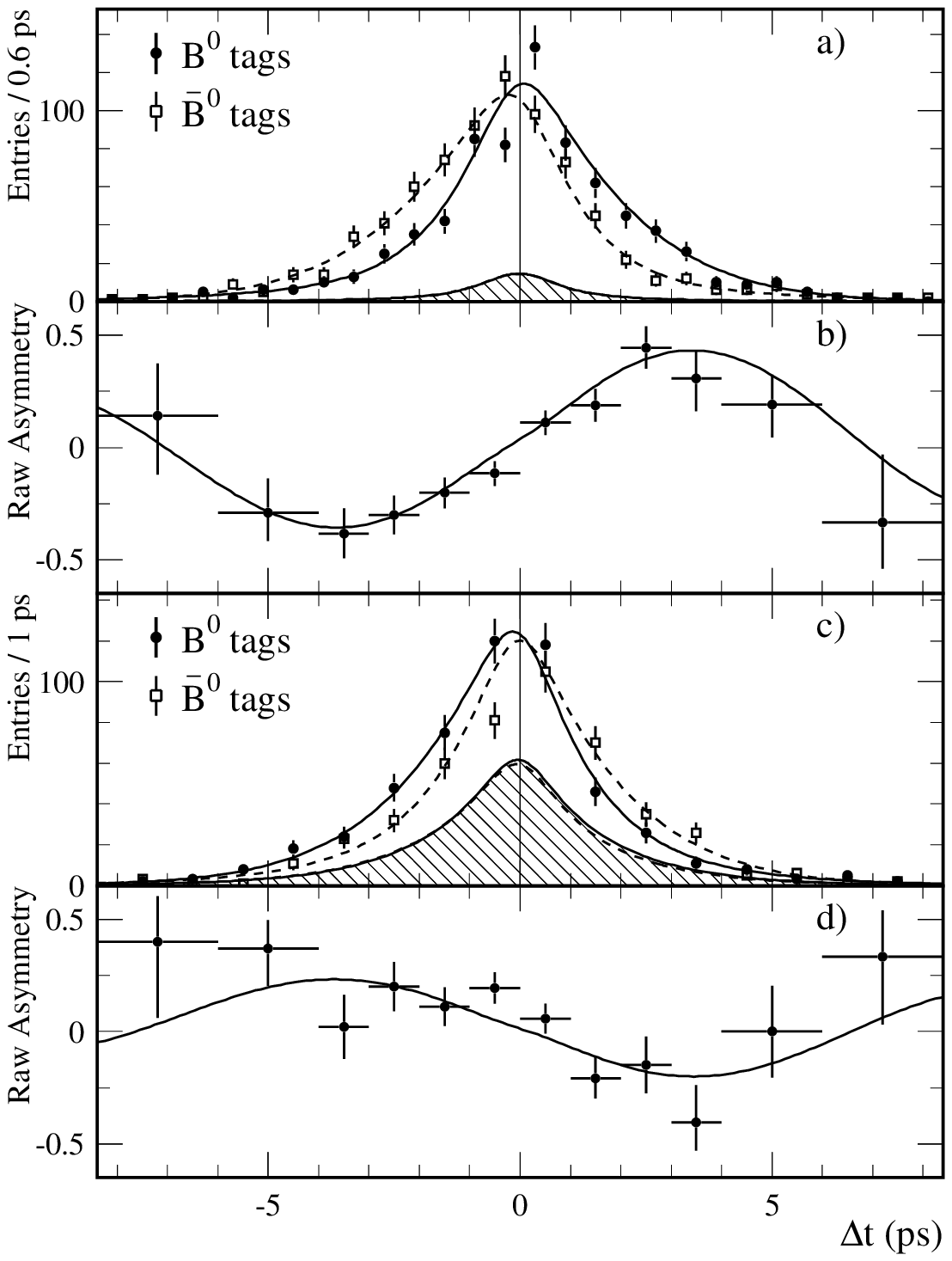}\quad
 \includegraphics[width=.48\linewidth]{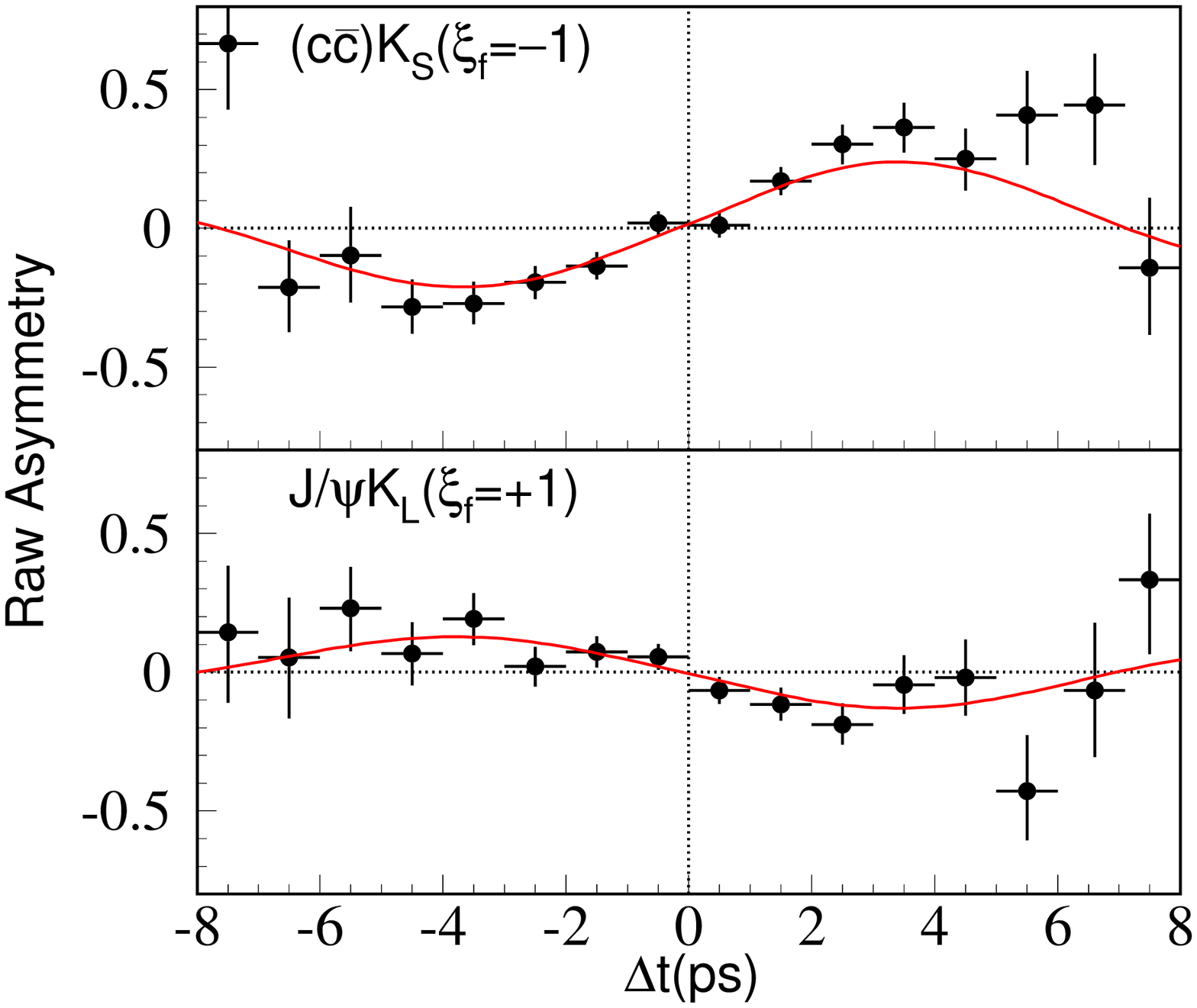}
}
\caption{\label{fig:dtdistros}Rate and flavor asymmetry vs $\Delta t$
 from BaBar (left) and Belle (right).  On the left appear
(a, c) the tagged-sample rates and (b, d) asymmetries for (a, b) $CP$-odd and
(c, d) $CP$-even modes.  On the right the asymmetries are given for (top) $CP$-odd ($\xi_f=\eta_f=-1$)  modes and (bottom) $CP$-even modes ($\xi_f=\eta_f=1$).  
}
\end{figure}

BaBar and Belle have measured $\sin 2\beta$ using a selection of events containing several charmionium $K^0_S$ modes as well as $J/\Psi K^0_L$ (where $\eta_f = +1)$. Figure~\ref{fig:ccKdistros} shows a list of charmonium modes used in the measurement. The data samples used for the measurement by the BaBar and Belle experiments are based on 82~fb$^{-1}$ and  140fb~$^{-1}$ respectively and the $\sin 2\beta$ result obtained by both experiments are show in Table~\ref{tab:s2b} whilst the $\Delta t$ distributions are shown in Figure~\ref{fig:dtdistros}.

\begin{table}[!ht]
\begin{center}
\begin{tabular}{|c|c|c|}
\hline
\multicolumn{1}{|c|}{\raisebox{-1.5ex}[0cm][0cm]{Measurements}}&
\multicolumn{2}{c|}{{Experiment}}\\
\hhline{|~--|}
&
BaBar&
Belle\\
\hline
$\sin 2\beta$&
0.741   $\pm$ 0.067  $\pm$ 0.034 &
0.733 $\pm$ 0.057 $\pm$ 0.028\\
\hline
\end{tabular}
\end{center}
\caption{\it The BaBar experiment measures $\sin 2 \beta$ using a 34-parameter likelihood fit to 2641 tagged events (which have a purity of 78\%. The Belle experiment use 3085 tagged events (with a purity of 76\%)}\label{tab:s2b}
\end{table}

The results obtained by both experiment are in agreement with each other and provide a new constraint on the upper vertex of the CKM triangle corresponding to $\beta$ whilst also being a clearly establishing $CP$ violation in $B$ meson decays.

\section{$\sin 2 \beta$ from $b \rightarrow sq\bar{q}$ decays}

There is also the possibility to observe $CP$ violation in $B$ channels which decay via  $b \rightarrow sq\bar{q}$ transitions with no $c$ quark in the final state. These processes are sensitive to the possibility of new physics, due to their smaller amplitudes interference terms being more significant and because of the possible contributions from virtual particles in the penguin loops such as SUSY particles. However, measurements for these decays are less accurate than those for $b \rightarrow sc\bar{c}$ decays due to the lower rates and higher backgrounds. Furthermore, the interpretation of the  measurements associated with  $b \rightarrow sq\bar{q}$ decays is harder, due to the simultaneous contributions from tree and penguin amplitudes. However the tree amplitude is CKM suppressed for the $b \rightarrow u$ transition such that the leading amplitudes tend to be $b \rightarrow s$ gluonic penguins. The best example for this is the decay $B \rightarrow \phi K^0_S$ where there is no tree contribution such that estimates of $\Delta \beta$ from ``tree pollution'' are as small as 0.01~\cite{bene}. 

The $B \rightarrow \phi K^0_S$ events are selected by reconstructing the $\phi$ from its $K^+ K^-$ decay and the  $K^0_S$ from $\pi^+\pi^-$ (BaBar also include $\pi^0 \pi^0$). In addition BaBar also reconstruct  $B \rightarrow \phi K^0_L$ decays, identifying the $K^0_L$ with the Instrumented Flux Return (IFR).

Using a dataset of 110~fb$^{-1}$ the BaBar experiment reconstructed 70 $\pm$ 9 $B \rightarrow \phi K^0_S$ events and 52 $\pm$ 16 $B \rightarrow \phi K^0_L$ events. From these events the results

\begin{center}
$S_{\phi K^0}$ = 0.47 $\pm$ 0.34 $^{+0.08}_{-0.06}$ \quad    $C_{\phi K^0}$ = 0.01 $\pm$ 0.33 $\pm$ 0.10
\end{center}

\noindent were obtained, were the first error is statistical and the second is systematic.

With a dataset of 140~fb$^{-1}$ Belle find  68 $\pm$ 11 events corresponding to a measurement of

\begin{center}
$S_{\phi K^0}$ = -0.96 $\pm$ 0.50 $^{+0.09}_{-0.11}       $    $C_{\phi K^0}$ = -0.15 $\pm$ 0.29 $\pm$ 0.07
\end{center}

\noindent were the first error is statistical and the second is systematic. The Belle measurement of $S_{\phi K^0}$ is 3.5 standard deviations from the standard model measurement of $\sin 2\beta$ obtained from the charmonium modes and could be evidence of new physics beyond the standard model. However there is also the BaBar measurement of $S_{\phi K^0}$ which is totally compatible with the standard model.

There is also the measurement of the non-resonant $B^0 \rightarrow K^+K^-K^0_S$, with events consistent with the $\phi$ mass in the $K^+K^-$ invariant mass plane rejected, performed by both BaBar and Belle. Measurements from BaBar and Belle indicate that the $B^0 \rightarrow K^+K^-K^0_S$ decay is in fact 104\% $\pm$ 20\% and 103 $\pm$ 15\% $CP$ even ($\xi_f$ = +1) respectively, and obtain the following results

\begin{eqnarray}
-\xi_fS_{K^+K^-K^0_S} &=& 0.56 \pm 0.25 \pm 0.04  \quad  C_{K^+K^-K^0_S} = -0.10 \pm 0.19 \pm 0.10 ~\mathrm{(BaBar)}\nonumber\\
-\xi_fS_{K^+K^-K^0_S} &=& 0.51 \pm 0.26 \pm 0.05 \quad   C_{K^+K^-K^0_S} = -0.17 \pm 0.16 \pm 0.04 ~\mathrm{(Belle)}\nonumber
\end{eqnarray}  

Both BaBar and Belle perform an analysis of the decay $B^0 \rightarrow \eta^\prime K^0_S$ which is reconstructed from $\eta^\prime \rightarrow \eta \pi^+\pi^-$ and $\eta^\prime \rightarrow\rho^0\gamma$. BaBar observe 203 $\pm$ 19 and Belle observe 244 $\pm$ 21 events leading to the measurements:

\begin{eqnarray}
S_{\eta^\prime K^0_S} &=& 0.02 \pm 0.34  \pm 0.03  \quad C_{\eta^\prime K^0_S} = 0.10 \pm 0.22 \pm 0.03 ~\mathrm{ (BaBar)}\nonumber \\
S_{\eta^\prime K^0_S} &=& 0.43 \pm 0.27 \pm 0.05 \quad   C_{\eta^\prime K^0_S} = 0.01 \pm 0.16 \pm 0.04 ~\mathrm{ (Belle)}\nonumber
\end{eqnarray}

The BaBar experiment has also looked at two new modes  $B^0 \rightarrow f_0K^0_S$ and $B^0 \rightarrow \pi^0K^0_S$ where the $f_0$ was reconstructed from its $\pi^+\pi^-$ decay. The decay vertex for  $B^0 \rightarrow \pi^0K^0_S$ was obtained by extrapolating the  $K^0_S$ back to the beam-spot and the x-y position of the beam-spot was used to constrain it.
Based on 111~fb$^{-1}$ and 110~fb$^{-1}$ the following results were obtained for $B^0 \rightarrow f_0K^0_S$ and $B^0 \rightarrow \pi^0K^0_S$ respectively:

\begin{eqnarray}
-\xi_fS_{f_0K^0_S} &=& 1.62 ^{+0.51}_{-0.56} \pm 0.10  \quad  C_{f_0K^0_S} = 0.27 \pm 0.36 \pm 0.12 \nonumber\\
-\xi_fS_{\pi^0K^0_S} &=& 0.48 ^{+0.38}_{-0.47} \pm 0.09  \quad   C_{\pi^0K^0_S} = 0.27 ^{+0.27}_{-0.28} \pm 0.06 \nonumber
\end{eqnarray}

\begin{figure}[!htbp]
\begin{center}
\scalebox{0.75}{%
 \includegraphics[width=.59\linewidth]{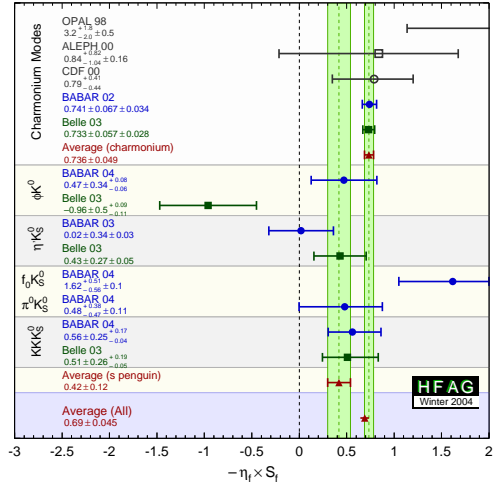}
 }
\caption{{\it Measurements of $\sin 2 \beta$ and their world averages.}}
\label{fig:sin2bsum}
\end{center}
\end{figure}

The measurement of $S_{f_0K^0_S}$ is 1.2 standard deviations from the physical limit and 1.7 standard deviations from the standard model whilst the measurement $-\xi_fS_{\pi^0K^0_S}$ is in total agreement with the standard model. All the modes relating to $\sin 2\beta$ discussed in this section are shown in Figure~\ref{fig:sin2bsum}.

\section{$B^0 \rightarrow {D^*}^{\mp} \pi^{\pm}$ and  $\sin (2\beta + \gamma$)}

The decay modes $B^0 \rightarrow {D^*}^{\mp} \pi^{\pm}$ have been
proposed for use in measurements of
$\sin(2\beta+\gamma)$~\cite{ref:book}, where $\beta = \arg{\left(-
V^{}_{cd} V_{cb}^* V^{}_{td} V_{tb}^*\right)}$ is well
measured (see Section~\ref{sec:sin2b}).

In the Standard Model the decays 
$B^0 \rightarrow D^{*+}\pi^-$ and $\overline{B}^0 \rightarrow D^{*+} \pi^-$
proceed through the $\overline{b} \rightarrow \overline{u}  c  
\overline{d}  $ and
$b \rightarrow c$ amplitudes $A_u$ and $A_c$. 
The relative weak phase between the two amplitudes 
in the usual Wolfenstein convention~\cite{ref:wolfen}
is $\gamma$.
When combined with $B^0-\overline{B}^0$  mixing, this yields a weak phase
difference of $2\beta+\gamma$ between the interfering amplitudes.
The decay rate is expressed by Equation~\ref{eq:tevo} where the amplitude of the sine term is proportional to

\begin{equation}
{\cal S}_{D^*\pi}=  (1-2\omega_{D^*\pi}) \, (s_t a_{D^*\pi} + s_m c_{D^*\pi}) 
        + s_t s_m b_{D^*\pi} (1-s_t \Delta\omega_{D^*\pi}),
\end{equation}

\noindent where $s_t = 1$ ($-1 $) when the tagged $B$ is identified as a $B^0(\overline{B}^0)$, 
and $s_m = 1$ ($-1 $) for ``unmixed'' (``mixed'') events.
The parameters $a_{D^*\pi}$, $b_{D^*\pi}$, and $c_{D^*\pi}$ are related to the physical parameters through:

\begin{eqnarray}
a_{D^*\pi}&\equiv& 2 r_{D^*\pi}\sin(2\beta+\gamma)\cos\delta' , \nonumber\\
b_{D^*\pi}&\equiv& 2 r^\prime\sin(2\beta+\gamma)\cos\delta_{D^*\pi} , \nonumber\\
c_{D^*\pi}&\equiv& 2\cos(2\beta+\gamma)(r_{D^*\pi}\sin\delta' -r^\prime\sin\delta_{D^*\pi}). 
\label{eq:abc}
\end{eqnarray}

Here $\delta_{D^*\pi}$ is the strong phase difference between $A_u$  and $A_c$ 
and $r = |A_u / A_c|$. Since $A_u$ is doubly CKM-suppressed with respect to $A_c$, one expects $r\sim 0.02$. Also $r'$ ($\delta^\prime$) is the effective magnitude of the ratio of amplitudes (strong phase difference) between the $b\rightarrow u \overline c d$ and $b\rightarrow c \overline u d$ amplitudes in the tagged $B$ decay. The parameters $a_{D^*\pi}$ and $c_{D^*\pi}$ have been measured by both the BaBar and Belle experiments and are shown in Table~\ref{tab:sin2bpg}.

\begin{table}[!ht]
\begin{center}
\begin{tabular}{|c|c|c|}
\hline
Experiment &
$a_{D^*\pi}$&
$c_{D^*\pi}$\\
\hline
BaBar&
-0.068  $\pm$ 0.038  $\pm$ 0.020&
-0.031 $\pm$ 0.070  $\pm$ 0.033\\
\hline
Belle&
-0.060  $\pm$ 0.040  $\pm$ 0.017 &
-0.049 $\pm$ 0.040 $\pm$ 0.019\\
\hline
\end{tabular}
\end{center}
\caption{{\it Measurements of $a_{D^*\pi}$ and $c_{D^*\pi}$ performed by the BaBar and Belle experiments. BaBar used their values to set the constraint $\sin(2\beta+\gamma) > 0.58$ at a 95\% confidence level.}}\label{tab:sin2bpg}
\end{table}

\section{Measurements of $V_{cb}$ and $V_{ub}$}

Given the semileptonic decay $B \rightarrow X\ell\nu$ one can reconstruct the neutrino where then it is possible to measure the hadronic mass $M_X$~\cite{mx}. In the case where $X = X_c$ the hadronic mass is greater than or equal to the mass of the $D$ meson. However when $X = X_u$ the hadronic mass distribution extends below the $D$ meson mass such that a cut $M_X < M^{cut}_X$ significantly enhances the relative contribution $b \rightarrow u$ compared to  $b \rightarrow c$. A measurement of the rate can then be used directly to measure $V_{cb}$ or $V_{ub}$. This approach was first used by CLEO~\cite{cleovxb} and is now being used by BaBar~\cite{babvxb} and Belle~\cite{bellevxb}.

\subsection{V{cb}}

We can study the moments of the invariant mass distributions, $<M_X>$ and $<M^2_X>$ where the moments are extracted directly from the measured $M_X$ and $M^2_X$ distributions directly, taking into account corrections for mass scales, the detector efficiency and small residual backgrounds. Theoretical calculations of the second moment $<M^2_X>$ have been performed using an Operator Product Expansion (OPE) in powers of the strong coupling constant $\alpha_s(m_b)$ and $1/m_b$ up to the order of ${\cal O}(1/m_b^3)$. A review on this subject can be found here~\cite{ope}. Combining measurements of $<M^2_X>$ for values of the momentum threshold with other measurements of semileptonic branching fractions and $B$ lifetimes we can further constrain the $b$ quark mass, $m_b$ in the kinematic mass scheme~\cite{mb} and the CKM matrix element $|V_{cb}|$~\cite{vcb}. Based on a dataset of 82~fb$^{-1}$ corresponding to 89 million $B$ meson pairs the following results where obtained.

\begin{center}$|V_{cb}|$ = (41.4 $\pm$ 0.4$_{exp}$ $\pm$ 0.4$_{HQE}$ $\pm$ 0.2$_{\alpha_S}$ $\pm$ 0.6$ _{\Gamma_{SL}}) \times 10^{-3}$

$m_b$ = 4.61 $\pm$ 0.05$_{exp}$ $\pm$ 0.04$_{HQE}$ $\pm$ 0.02$_{\alpha_S}$ GeV/c$^2$
\end{center}
 
\noindent where in both cases the first error is due to the experimental uncertainty of the measurement, the second is a consequence of Heavy Quark Expansion (HQE) and the third is due to the uncertainty contribution from $\alpha_S$. For $|V_{cb}|$ the last error is due to the uncertainty from the HQE of the semileptonic width $\Gamma_{SL}$.  
\subsection{V{ub}}

BaBar use a data sample of 82~fb$^{-1}$ containing 89 million $B\bar{B}$ pairs. One $B$ is fully reconstructed and the other $B$ is analyzed for semileptonic decays. For semileptonic decay candidates, a high momentum lepton is required such that $P >$ 1.0~GeV/c. All the reconstructed particles not associated with the fully reconstructed $B$ are used to calculate the missing four-momentum of the semileptonic decay. This missing four-momentum is taken as being the neutrino four-momentum. A kinematic fit is performed on the event to determine $M_X$. Additional cuts are then applied to improve the quality of events, such as requiring the invariant mass of the neutrino to be small in order to improve the $M_X$ resolution. A fit is performed on the $M_X$ distribution whereby the region $M_X >$ 1.55~GeV/c$^2$ is dominated by $B \rightarrow X_c\ell\nu$ and is used to fix the scale for the $b \rightarrow c$ distribution in $M_X$. The number of $B \rightarrow X_u\ell\nu$ events below  $M_X <$  1.55~GeV/c$^2$ is then extracted from the fit from which they calculate:

\begin{center}
$|V_{ub}|$  = (4.62 $\pm$ 0.28 $\pm$ 0.27 $\pm$ 0.40 $\pm$ 0.26) $\times 10^{-3}$  
\end{center}

\noindent where the first error is statistical, the second is systematic, the third is due to the error on $B \rightarrow X_u\ell\nu$ modeling and the fourth is due to the error on the known relationship between the rate and $|V_{ub}|$.

Belle follow a similar procedure to that descried above. They require one fully reconstructed $B$ in the events and look for the other $B$ to decay semileptonically. With the knowledge of which particles belong to the fully reconstructed $B$ and which belong to the semileptonic decay, the values for $M_X$ and $q^2$ (mass of the virtual $W$ squared) are calculated. Candidates are required to have $M_X < $ 1.5~GeV/c$^2$ and  $q^2 > $ 7~GeV/c$^2$. The $b \rightarrow c$ events are subtracted and from the $B \rightarrow X_u\ell\nu$ events in the $M_X$ distribution they find

\begin{center}
$|V_{ub}|$  = (4.66 $\pm$ 0.28 $\pm$ 0.35 $\pm$ 0.58 $\pm$ 0.17 $\pm$ 0.08) $\times 10^{-3}$  
\end{center}

\noindent where the first error is statistical, the second is systematic, the third is theoretical and the fourth and fifth errors are due to $b \rightarrow c$ and $b \rightarrow u$ modeling respectively.

\section{Summary}
\begin{figure}[!htbp]
\begin{center}
\scalebox{0.80}{%
 \includegraphics[width=.59\linewidth]{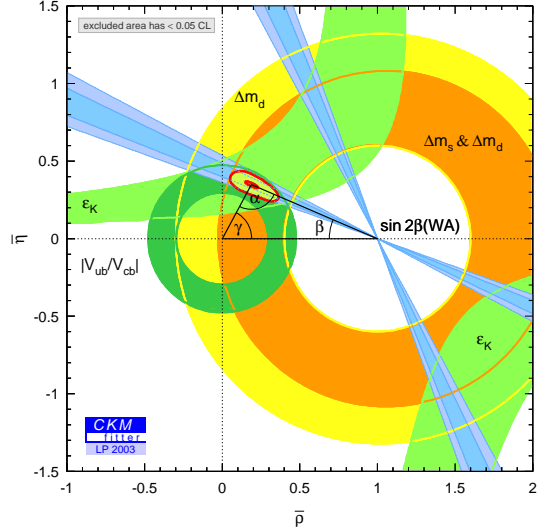}
 }
\caption{{\it Measurements of $\sin 2 \beta$ from charmonium modes compared in the $\bar{\rho}-\bar{\eta}$ plane. Where $\rho$ and $\eta$ are the parameters of the Wolfenstein parameterization of the CKM matrix.}}
\label{fig:tri}
\end{center}
\end{figure}

BaBar and Belle have established $CP$ violation in $B^0$ decays through the measurement of $\sin 2 \beta$, where the constraint on the CKM unitarity triangle is shown in Figure~\ref{fig:tri}. This $CP$ asymmetry  is well accommodated within the standard model, however we are starting to see evidence for inconsistencies from charmless decays. Also, $CP$ violation in the decay $B^0 \rightarrow \pi^+ \pi^-$ has been reported by Belle, though not yet confirmed by BaBar. We look forward to additional data which will provide definitive results regarding these measurement, helping to further constrain the angles $\alpha$ and $\gamma$ and with the help of theorists provide more precise measurements of the CKM elements $V_{cb}$ and $V_{ub}$.

\section{Acknowledgements}
I am extremely grateful to the organizers of the conference and would like to thank them for providing an interesting and enlightening physics program. It is also a pleasure to thank the BaBar and Belle experiments for their achievements in obtaining high luminosity with which they have provided a wide range of physics results.

\end{document}